# Synchro-Transient-Extracting Transform for the Analysis of Signals with Both Harmonic and Impulsive Components

Yunlong Ma, Gang Yu, Tianran Lin and Qingtang Jiang

*Abstract*—Time-frequency analysis (TFA) techniques play an important role in the field of machine fault diagnosis attributing to their superiority in dealing with nonstationary signals. Synchroextracting transform (SET) and transient-extracting transform (TET) are two newly emerging techniques that can produce energy concentrated representation for nonstationary signals. However, SET and TET are only suitable for processing harmonic signals and impulsive signals, respectively. This poses a challenge for each of these two techniques when a signal contains both harmonic and impulsive components. In this paper, we propose a new TFA technique to solve this problem. The technique first divides time-frequency (TF) coefficients into two types according to the chirp rate of the signal, and then TF ridges are respectively extracted to prevent the limitation for the independent use of SET and TET. This allows the technique to generate an energy-concentrated TF representation for both harmonics and impulses in the signal. Furthermore, we theoretically demonstrate that the proposed technique retains the signal reconstruction capability. The effectiveness of the proposed technique is verified using numerical and real-world signals.

*Index Terms*—time-frequency analysis, synchroextracting transform , transient-extracting transform , synchro-tran-sient-extracting transform.

## I. INTRODUCTION

MANY real-world signals, such as gravitational waves or ultrasonic waves from bats, demonstrate nonstationary characteristics. The time-frequency (TF) analysis (TFA) technique is a powerful tool to extract features from nonstationary signals [1],[2]. Traditional TFA techniques include the short-time Fourier transform (STFT) [3] and continuous wavelet transform [4]. Unfortunately, the time-frequency representation (TFR) of these techniques is blurry due to the Heisenberg uncertainty principle. To overcome this shortcoming, many postprocessing tools have been developed.

The reassignment (RS) [5] was an early proposed postprocessing technique that performs a two-dimensional reassignment of the TF spectrogram by computing the instantaneous frequency (IF) estimate and group delay (GD) estimate of a signal. However, the RS cannot reconstruct the original signal. The synchrosqueezing transform (SST) [6] squeezes the TF coefficients into the IF estimate, which allows to obtain an energy-concentrated TFR while retaining the ability to reconstruct the original signal. [7], [8]. However, SST can only produce concentrated TFR for slowly time-varying signals, while the TFR for strong time-varying signals is blurry [9]-[12]. Furthermore, the SST is less effective in dealing with signals contaminated by noise [13]. This is because SST may confuse the IF trajectory of the noise with the IF trajectory of the signal, which would lead to incorrect reassignment. Yu et al. proposed a synchroextracting transform (SET) [14] that only retains the TF coefficients on the IF trajectory to overcome the drawbacks of SST. However, SET and SST use the same IF estimate, which leads to an unsatisfactory characterization of SET for strong time-varying signals [15].

The slowly time-varying signals are also called harmonic-like signals, and such signals are characterized by a very small chirp rate. In addition to harmonic-like signals, another class of signals called impulsive-like signals is widely available in the real world [16]. The IF of the impulsive-like signal is almost perpendicular to the time axis in the TF plane, which leads to a very large chirp rate of the signal, and thus the SST and SET cannot obtain a concentrated TFR. Although the IF of impulsive-like signals varies drastically with time, their GD varies slowly with frequency. This inspires us to deal with impulsive-like signals in the frequency domain. He et al. proposed a time-reassigned synchrosqueezing transform (TSST) [17] that reassigns the TF coefficients along the time axis by calculating the GD estimate of a signal. The disadvantages of TSST are its inability to generate energy-concentrated TFR for harmonic-like signals and its heavy influence by noise. Yu proposed a transient-extracting transform (TET) [18] that only extracts the TF coefficients on the GD trajectory to overcome these disadvantages. Since the TET and TSST use the same GD estimate, the TFR of the TET is unsatisfactory for harmonic-like signals [19].

As mentioned above, SET is suitable for dealing with harmonic-like signals and TET is suitable for impulsive-like signals. Therefore, when both harmonic-like and impulsive-like components are present within a signal, a single technique cannot generate an energy-concentrated TFR for both components. However, real signals, such as gear fault signals [20] and rub-impact fault signals, often contain multiple components, such as harmonic, impulsive and frequency modulation (FM) components. If such fault signals cannot be detected in time, they can lead to production downtime and

This work was financially supported in part by the National Natural Science Foundation of China under Grant 62271230.
Y. Ma and G. Yu are with the School of Electrical Engineering, University of Jinan, Jinan 250022 , China (e-mail:202121200832@stu.ujn.cn; cse_yug@ujn.edu.cn).
T. Lin is with the School of Mechanical and Automatic Engineering, Qingdao University of Technology, Qingdao 266520, China.
Q. Jiang is with the Department of Mathematics and Statistics, University of Missouri-St. Louis, St. Louis, MO 63132 USA.

economic losses, and can even lead to major safety accidents. In this paper, we propose a novel TF postprocessing technique to solve the above problems. The technique is divided into three steps. First, the STFT of a signal is divided into two parts, i.e., harmonic-like structure and impulsive-like structure, based on the chirp rate of the signal. Then, the TF ridges of the two structures are extracted separately using different techniques. Finally, the extracted results are summed up to obtain an energy-concentrated TFR. Therefore, the proposed technique can simultaneously provide energy-concentrated TFRs for the harmonic, impulsive and FM components of the signal. In addition, the technique can reconstruct the original signal and shows good noise robustness in the analysis of nonstationary signals. The rest of this paper is organized as follows. In Section II, we review some fundamental notations and definitions. Section III introduces our proposed technique. The numerical and experimental validations are given in Section IV and Section V, respectively. The conclusion is drawn in Section VI.

## II. A REVIEW OF KNOWLEDGE

### A. Short-time Fourier Transform

The STFT of a signal $s \in L^2(R)$ with respect to the real and even window $g \in L^2(R)$ is defined as

$$V(t,\omega) = \int_{-\infty}^{\infty} s(u)g(u-t)e^{-i\omega(u-t)}du \quad (1)$$

where $s(u)$ is the analyzed signal and $i$ is the imaginary unit. In mechanical systems, harmonics, impulses and FM signals are three common types of signals, and STFT is a classical TFA technique for representing these three types of signals. However, since the window has a length, the TF energy of the STFT is diffused inside the window. Herein, we use a numerical signal and Fig. 1 provides the results of STFT, SET and TET for the signal. Compared with the ideal TFR, the TF resolution of the STFT result cannot be arbitrarily small in the time and frequency axis. The limitation of STFT have given a rise to the development of TF postprocessing techniques.

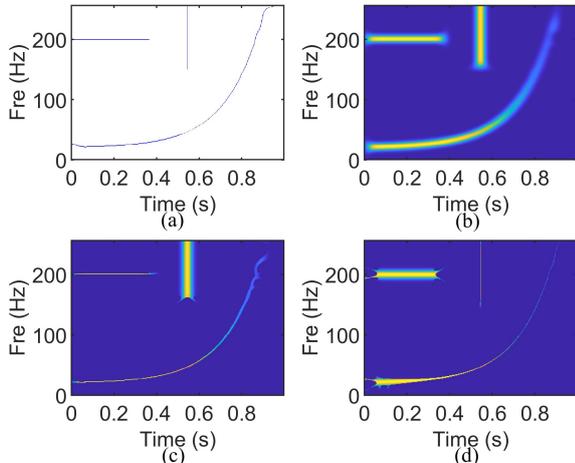

Fig. 1. (a) Ideal TFR, (b) STFT result, (c) SET result, and (d) TET result.

### B. Synchroextracting Transform

To achieve a sharper TFR, enhancing energy concentration has become an important research direction, in which SET is an effective technique. The key idea of SET is to extract the TF coefficients on the IF trajectory and remove other TF coefficients. The SET is defined as

$$S_e(t,\omega) = V(t,\omega)\delta(\omega - \hat{\omega}(t,\omega)) \quad (2)$$

where $\hat{\omega}(t,\omega)$ is the IF estimator, which can be calculated from the following:

$$\hat{\omega}(t,\omega) = \mathrm{Re}\left(\frac{\partial_t V(t,\omega)}{iV(t,\omega)}\right) \quad (3)$$

where $\mathrm{Re}(\cdot)$ represents the real part of a complex number. The SET can generate energy concentrated TFR for harmonic signals but not for impulsive signals. Fig. 1(c) shows that the SET can yield a clear TFR for the harmonic component but no improvement over STFT for the impulsive component. Then, the TET technique is proposed for processing impulsive signals.

### C. Transient-extracting Transform

The key idea of TET is to retain the TF coefficients on the GD trajectory and remove other TF coefficients. The TET is formulated as

$$T_e(t,\omega) = V(t,\omega)\delta(t - \hat{t}(t,\omega)) \quad (4)$$

where

$$\hat{t}(t,\omega) = t - \mathrm{Im}\left(\frac{\partial_\omega V(t,\omega)}{V(t,\omega)}\right) \quad (5)$$

where $\mathrm{Im}(\cdot)$ represents the imaginary part of a complex number. The TET can generate energy-concentrated TFR for impulsive signals but not for harmonic signals. Fig. 1(d) shows that the TET can yield a clear TFR for the impulsive component but no improvement over STFT for the harmonic component.

### D. A Performance Evaluation of SET and TET in the Analysis of FM Signals

The previous sections introduced the advantages and disadvantages of SET and TET. In this section, we use an FM signal to further analyze the property of SET and TET. In general, an amplitude-modulated frequency-modulated signal can be described as $s(t) = A(t)e^{i\varphi(t)}$, where $\varphi'(t)$ is the IF of the signal. Herein, we use an FM signal as follows,

$$s_f(t) = Ae^{i(a+bt+0.5ct^2)} \quad (6)$$

where $b+ct$ and $c$ are the IF trajectory and the chirp rate of the signal, respectively. Assume that the window function has the following form,

$$g(t) = e^{-(2\beta)^{-1}t^2}. \quad (7)$$

Then, substituting (6) into (3) to have

$$\hat{\omega}(t,\omega) = b + ct + c^2\beta^2(\omega - b - ct)(1+c^2\beta^2)^{-1}. \quad (8)$$

It is shown that $\hat{\omega}(t,\omega)$ cannot accurately locate the IF trajectory of the FM signal. According to (8), we can obtain the error between the IF estimator and the IF trajectory as

$$r_{\hat{\omega}}(t,\omega) = |\hat{\omega}(t,\omega) - (b+ct)| = |\omega - b - ct|(1+c^{-2}\beta^{-2})^{-1}. \quad (9)$$

When the signal is a harmonic, $r_{\hat{\omega}}(t,\omega) = 0$, which means that $\hat{\omega}(t,\omega)$ can accurately locate the IF trajectory and therefore SET can generate an ideal TFR. When the chirp rate increases, $r_{\hat{\omega}}(t,\omega)$ increases, which leads to an increase in the distance between $\hat{\omega}(t,\omega)$ and the IF trajectory, and the TFR of the SET will get blurred. Therefore, it can be inferred that the concentration of the SET result decreases with the increase of chirp rate.

Next, we illustrate the relationship between the performance of the TET and the chirp rate. The signal $s(t) = A(t)e^{i\varphi(t)}$ can be represented in the frequency domain as $\hat{s}(\omega) = A(\omega)e^{i\varphi(\omega)}$, where $-\varphi'(\omega)$ is the GD of the signal. The signal described by (6) can be rewritten as

$$\hat{s}_f(\omega) = A\sqrt{2\pi \cdot c^{-1}} e^{i(a + \frac{\pi}{4} - \frac{(\omega-b)^2}{2c})} \quad (10)$$

where $(\omega-b)/c$ is the GD trajectory of the signal. Based on (5), we can obtain

$$\hat{t}(t,\omega) = \frac{c^2\beta^2}{1+c^2\beta^2} \cdot \frac{\omega-b}{c} + \frac{t}{1+c^2\beta^2}. \quad (11)$$

It can be observed that $\hat{t}(t,\omega)$ cannot accurately locate the GD trajectory of the FM signal. According to (11), we can obtain the error between the GD estimator and the GD trajectory as

$$r_{\hat{t}}(t,\omega) = \left|\hat{t}(t,\omega) - (\omega-b)/c\right| = \left|\frac{t - (\omega-b)/c}{1+c^2\beta^2}\right|. \quad (12)$$

When the chirp rate is small, $r_{\hat{t}}(t,\omega)$ is large, which indicates that the distance between $\hat{t}(t,\omega)$ and the GD trajectory is large and therefore the TFR of TET is blurry. As the chirp rate increases, $r_{\hat{t}}(t,\omega)$ decreases, which means that the distance between $\hat{t}(t,\omega)$ and the GD trajectory decreases and the concentration of the TFR of TET becomes higher. Therefore, it can be inferred that the concentration of the TET result increases with the increase of chirp rate.

## III. SYNCHRO-TRANSIENT-EXTRACTING TRANSFORM (STET) AND IMPROVED STET

### A. The Basic Idea of STET

The above analysis shows that the concentration of SET and TET vary with the chirp rate. It should be noted that the change in the concentration of the two techniques with the chirp rate is opposite, as shown in Fig. 2, where the white numbers are the Rényi entropy. The Rényi entropy is an indicator to evaluate the energy concentration of a TFR [21]. In general, the more concentrated the TFR is, the lower the Rényi entropy is. It can be observed from Fig. 2 that as the chirp rate increases, the SET results gradually diverge, while the TET results gradually concentrate. Therefore, when a signal contains both small chirp components and large chirp components, a single technique cannot generate a concentrated TFR for each component. From Fig. 2, there exists a boundary at which SET and TET have the same concentration. On one side of the boundary, SET has a higher concentration, and on the other side, TET is better. If this boundary could be found, this would provide a selection strategy for the use of SET and TET. In this paper, a novel TFA technique is proposed by finding the boundary.

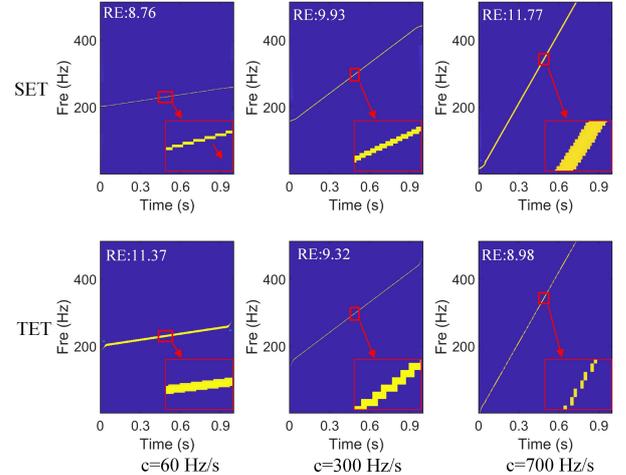

Fig. 2. SET results and TET results at different chirp rates and their Rényi entropy (white numbers).

The proposed technique is to find a chirp rate boundary that can differentiate the performance of SET and TET. By comparing the relationship between the chirp rate of an analyzed signal and the boundary, the technique with higher concentration in SET and TET is used to process the signal. From the previous section, $r_{\hat{\omega}}(t,\omega)$ and $r_{\hat{t}}(t,\omega)$ reflect the concentration of SET and TET, respectively. Therefore, $r_{\hat{\omega}}(t,\omega)$ and $r_{\hat{t}}(t,\omega)$ can be used as the criteria for selecting SET or TET. For instance, if $r_{\hat{\omega}}(t,\omega)$ is smaller than $r_{\hat{t}}(t,\omega)$, the difference between the IF estimator and the IF trajectory is smaller than the difference between the GD estimator and the GD trajectory, which results in the result obtained using SET will be closer to the TF trajectory of the signal than the result obtained using TET. Therefore, when $r_{\hat{\omega}}(t,\omega)$ is smaller than $r_{\hat{t}}(t,\omega)$, the signal is more suitable to be processed by SET and vice versa by TET. This means that the chirp rate when $r_{\hat{\omega}}(t,\omega) = r_{\hat{t}}(t,\omega)$ can be used as the boundary as follows:

$$|c| = \beta^{-2/3}. \quad (13)$$

In practice, for the FM signal described by (6), we do not know its chirp rate in advance. To this regard, we propose a chirp rate estimator. Based on (8) and (11), we have

$$\partial_\omega \hat{\omega}(t,\omega) = \beta^2 c^2(1+\beta^2 c^2)^{-1}, \partial_t \hat{\omega}(t,\omega) = c(1+\beta^2 c^2)^{-1} \\ \partial_\omega \hat{t}(t,\omega) = \beta^2 c(1+\beta^2 c^2)^{-1}, \quad \partial_t \hat{t}(t,\omega) = (1+\beta^2 c^2)^{-1} \quad (14)$$

where $\partial_t(\cdot)$ and $\partial_\omega(\cdot)$ denote the partial derivatives with respect to $t$ and the partial derivatives with respect to $\omega$, respectively. Based on (14), the chirp rate estimator can be calculated as

$$\hat{c}(t,\omega) = \begin{cases} \partial_t \hat{\omega}(t,\omega)/\partial_t \hat{t}(t,\omega), & \text{if } \partial_t \hat{t}(t,\omega) \neq 0 \\ \text{Inf}, & \text{if } \partial_t \hat{t}(t,\omega) = 0 \end{cases}. \quad (15)$$

When the chirp rate of a signal is less than or equal to the boundary value, SET is used to process the signal, otherwise TET is used. Finally, a high-resolution result is obtained by adding the result of SET and the result of TET. Based on this idea, a novel technique, termed as the synchro-transient-extracting transform (STET) can be obtained as

$$S_e(t,\omega) = \begin{cases} V(t,\omega)\delta(\omega-\hat{\omega}(t,\omega)), & \text{if } |\hat{c}(t,\omega)| \leq \beta^{-2/3} \\ 0, & \text{otherwise} \end{cases}$$

$$T_e(t,\omega) = \begin{cases} V(t,\omega)\delta(t-\hat{t}(t,\omega)), & \text{if } |\hat{c}(t,\omega)| > \beta^{-2/3} \\ 0, & \text{otherwise} \end{cases} \quad (16)$$

$$S(t,\omega) = S_e(t,\omega) + T_e(t,\omega).$$

Fig. 3 shows the STET result for the numerical signal in Fig. 1. It is shown that the proposed technique can yield a highly concentrated TFR for harmonic, impulsive and FM components. This indicates that STET breaks the restrictions on the signal model in SET and TET, and that STET combines the advantages of SET and TET.

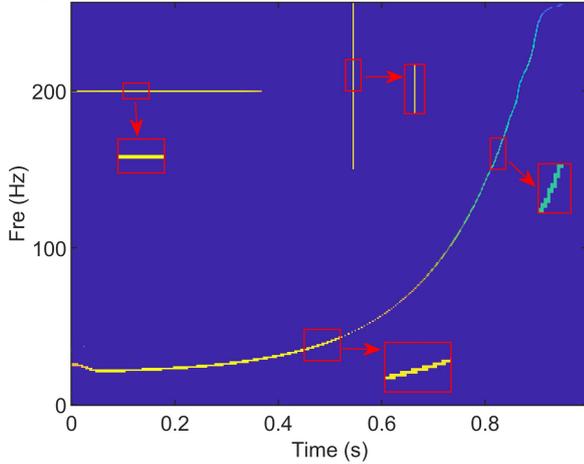

Fig. 3. STET result.

### B. An Improved Algorithm of STET

From the previous section, it is clear that the result of STET is the sum of the result of SET and the result of TET. However, SET and TET are derived based on purely harmonic and purely impulsive signals, respectively. When the signals within the window cannot be approximated by pure harmonics and pure impulses, $\hat{\omega}(t,\omega)$ and $\hat{t}(t,\omega)$ cannot accurately locate the IF and GD, which results in the result of SET and the result of TET will be blurred. As shown in Fig. 3, although STET can generate energy concentrated TFRs for the harmonic and impulsive components, there are still some energies around the TF trajectory for the FM component, which makes it cannot accurately interpret the TF features of the FM component. Therefore, we improve the SET and TET in STET so that STET can accurately locate the TF trajectory of FM signals.

*1) Improved SET*

For the FM signal described by (6), the goal of the improved SET is to improve the IF estimator $\hat{\omega}(t,\omega)$ so that it can accurately locate the IF trajectory $b+ct$. It can be found form (8) that $\hat{\omega}(t,\omega)$ cannot accurately locate the IF $b+ct$ of the FM signal due to the remainder $c^2\beta^2(\omega-b-ct)(1+c^2\beta^2)^{-1}$. Therefore, the goal of the improved IF estimator is to eliminate the remainder $c^2\beta^2(\omega-b-ct)(1+c^2\beta^2)^{-1}$. According to (11) and (14), we can get:

$$\begin{aligned}&(\hat{t}(t,\omega)-t)\partial_t\hat{\omega}(t,\omega)/\partial_t\hat{t}(t,\omega)\\&=c^2\beta^2(\omega-b-ct)(1+c^2\beta^2)^{-1}.\end{aligned} \quad (17)$$

According to (8) and (17), the improved IF estimator $\hat{\omega}^{[2]}(t,\omega)$ can be written as:

$$\hat{\omega}^{[2]}(t,\omega) = \hat{\omega}(t,\omega) - (\hat{t}(t,\omega)-t)\partial_t\hat{\omega}(t,\omega)/\partial_t\hat{t}(t,\omega). \quad (18)$$

Using (18), we can obtain the IF estimate of (6) as,

$$\hat{\omega}^{[2]}(t,\omega) = b+ct. \quad (19)$$

In addition to IF estimate, it is necessary to rectify the amplitude of the STFT on the IF trajectory. We first calculate the STFT of (6) on the IF trajectory as follows:

$$V(t,b+ct) = s_f(t)\sqrt{2\pi\beta(1-i\beta c)^{-1}}. \quad (20)$$

To make the TF coefficients on the IF trajectory only reflect the content of the signal $s_f(t)$, we construct a new STFT as follows:

$$E(t,\omega) = V(t,\omega)\sqrt{(1-i\beta c)(2\pi\beta)^{-1}}. \quad (21)$$

Finally, the improved SET is given by

$$S_e^{[2]}(t,\omega) = E(t,\omega)\delta(\omega-\hat{\omega}^{[2]}(t,\omega)). \quad (22)$$

*2) Improved TET*

For the FM signal described by (10), (11) shows that the conventional GD estimator cannot accurately estimate the GD trajectory $(\omega-b)/c$ of the FM signal. The goal of the improved TET is to improve the GD estimator so that it can accurately locate the GD trajectory. According to (8), (11) and (14), we can obtain:

$$\begin{aligned}&\frac{\partial_\omega \hat{t}(t,\omega)}{\partial_\omega \hat{\omega}(t,\omega)}(\omega-\hat{\omega}(t,\omega))+\hat{t}(t,\omega)\\&=\frac{1}{c}(\frac{c^2\beta^2(\omega-b)+ct-c^2\beta^2(\omega-b-ct)}{1+c^2\beta^2}+\omega-b-ct) \quad (23)\\&=\frac{\omega-b}{c}.\end{aligned}$$

It can be seen that (23) can accurately locate the GD trajectory of the FM signal. Therefore, the improved GD $\hat{t}^{[2]}(t,\omega)$ estimator is defined as:

$$\hat{t}^{[2]}(t,\omega) = \frac{\partial_\omega \hat{t}(t,\omega)}{\partial_\omega \hat{\omega}(t,\omega)}(\omega-\hat{\omega}(t,\omega))+\hat{t}(t,\omega). \quad (24)$$

Similarly, we rectify the magnitude of the STFT on the GD trajectory. The STFT of (10) on the GD trajectory as follows:

$$V(c^{-1}(\omega-b),\omega) = \hat{s}_f(\omega)\sqrt{\beta c(i+c\beta)^{-1}}e^{-i\frac{\omega b-\omega^2}{c}}. \quad (25)$$

We construct a new STFT as follows:

$$F(t,\omega) = V(t,\omega)\sqrt{(\beta c)^{-1}(i+\beta c)}e^{i\frac{\omega b-\omega^2}{c}} \quad (26)$$

where $b$ can be calculated from the following equation:

$$b = \omega - \hat{t}^{[2]}(t,\omega)(\partial_\omega \hat{\omega}(t,\omega)/\partial_\omega \hat{t}(t,\omega)). \quad (27)$$

Finally, the improved TET is given by

$$T_e^{[2]}(t,\omega) = F(t,\omega)\delta(t - \hat{t}^{[2]}(t,\omega)). \tag{28}$$

Based on (22) and (28), the improved STET is denoted as:

$$S_e^{[2]}(t,\omega) = \begin{cases} E(t,\omega)\delta(\omega - \hat{\omega}^{[2]}(t,\omega)), & \text{if } |\hat{c}(t,\omega)| \le \beta^{-2/3} \\ 0, & \text{otherwise} \end{cases}$$

$$T_e^{[2]}(t,\omega) \tag{29}$$

$$= \begin{cases} F(t,\omega)\delta(t - \hat{t}^{[2]}(t,\omega)), & \text{if } |\hat{c}(t,\omega)| > \beta^{-2/3} \\ 0, & \text{otherwise} \end{cases}$$

$$S_I(t,\omega) = S_e^{[2]}(t,\omega) + T_e^{[2]}(t,\omega).$$

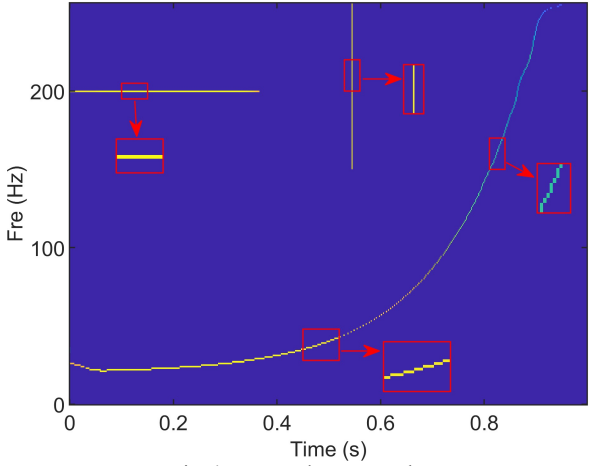

Fig. 4. Improved STET result.

Fig. 4 shows the improved STET result for the numerical signal in Fig. 1. Compared to STET, the energy concentration of the improved STET is further improved, and the energy only exists on the TF trajectory.

### C. Signal Reconstruction

Since the improved STET result is the sum of the improved SET result and the improved TET result, the reconstructed signal can be obtained by reconstructing the improved SET result and the improved TET result separately and then adding them together. According to (22), the FM signal that is processed by the improved SET can be reconstructed by the following equation:

$$\begin{aligned} S_e^{[2]}(t,\omega)\big|_{\omega-(b+ct)=0} &= E(t,\omega)\big|_{\omega-(b+ct)=0} \\ &= V(t,b+ct)\sqrt{(1-i\beta c)(2\pi\beta)^{-1}} \\ &= s_f(t). \end{aligned} \tag{30}$$

According to (28), the FM signal that is processed by the improved TET can be reconstructed by the following equation:

$$\begin{aligned} &\frac{1}{2\pi}\int_{-\infty}^{\infty} T_e^{[2]}(\hat{t}^{[2]}(t,\omega),\omega)e^{i\omega t}d\omega \\ &= \frac{1}{2\pi}\int_{-\infty}^{\infty} V(c^{-1}(\omega-b),\omega)\sqrt{(\beta c)^{-1}(i+\beta c)}e^{i\frac{\omega b-\omega^2}{c}}e^{i\omega t}d\omega \\ &= \frac{1}{2\pi}\int_{-\infty}^{\infty} \hat{s}_f(\omega)e^{i\omega t}d\omega \\ &= s_f(t). \end{aligned} \tag{31}$$

Therefore, the improved STET can reconstruct the original signal by the following expression:

$$\begin{aligned} s_1(t) &= S_e^{[2]}(t,\omega)\big|_{\omega-(b+ct)=0} \\ s_2(t) &= \frac{1}{2\pi}\int_{-\infty}^{\infty} T_e^{[2]}(\hat{t}^{[2]}(t,\omega),\omega)e^{i\omega t}d\omega \\ s(t) &= s_1(t) + s_2(t). \end{aligned} \tag{32}$$

## IV. NUMERICAL VALIDATION

In this section, a numerical signal is used to illustrate the advantages of the improved STET. Fig. 5 shows the TF information of the numerical signal, which is composed of three Hermite functions and a strongly frequency modulated component. The TFR of the signal shows the shape of a traditional Tai-Chi symbol.

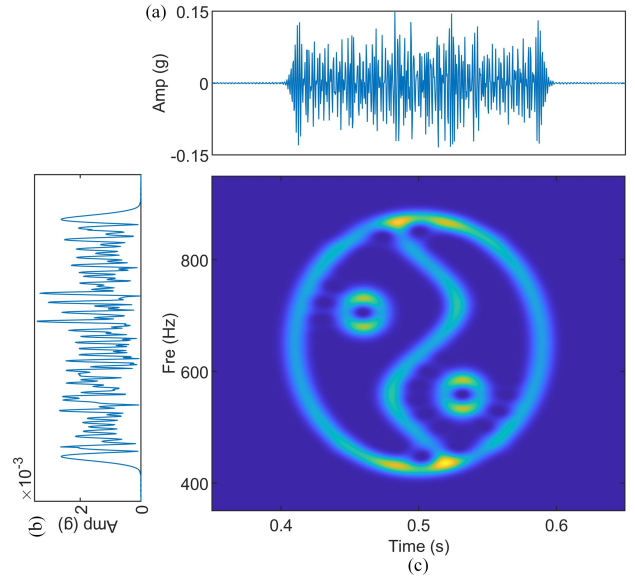

Fig. 5. (a) Waveform of the real part of the signal, (b) spectrum of the signal, and (c) STFT result.

The TFR in Fig. 5 is the STFT result, which suffers from a low TF resolution. For comparison, the TFRs of different techniques are shown in Fig. 6 and Fig. 7, where the white numbers are the Rényi entropy. It is shown that both SET and TET are unable to generate an energy-concentrated TFR. The concentration of the RS result is significantly higher than that of SET and TET, which is due to the fact that RS performs a two-dimensional reassignment for the spectrogram. However, there are still some energies around the TF trajectories. The STET result are more concentrated than the above techniques, but the limitation of the first-order estimator makes it extract some TF coefficients that are not on the TF trajectory. This problem is solved in the result of the improved STET. It can be observed that the improved STET result has a high energy concentration and it is further seen in the zoomed version that the improved STET generates a more concentrated TFR than the rest of the techniques. From the Rényi entropy results, the improved STET has the smallest Rényi entropy, which quantitatively indicates that its energy concentration is higher than other techniques. Figs. 7(e) and (f) shows the classification results of the improved STET. The results show that the improved STET reasonably distinguishes the harmonic-like structure and the impulsive-like structure of the signal, which leads to an energy-concentrated result for both parts.

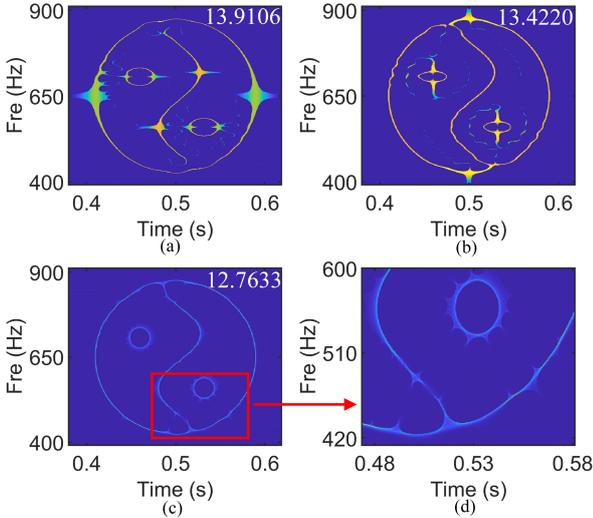

Fig. 6. (a) SET result, (b) TET result, (c) RS result, (d) zoom of the RS result.

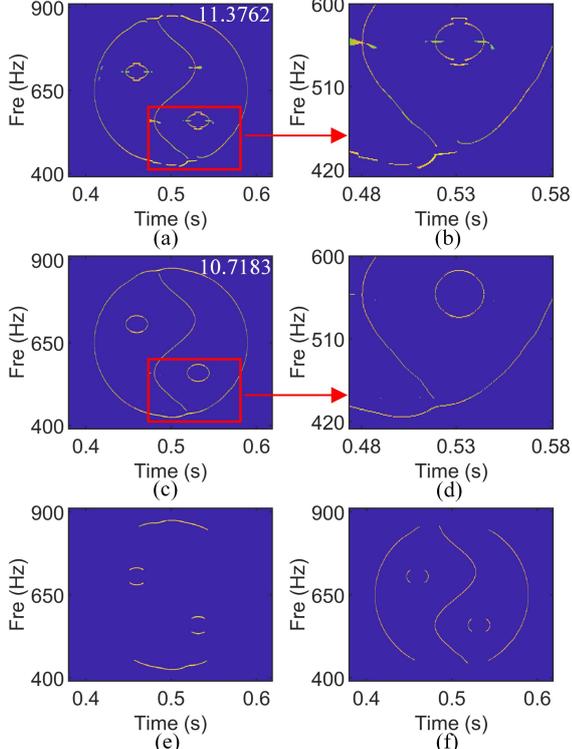

Fig. 7. (a) STET result, (b) zoom of the STET result, (c) improved STET result, (d) zoom of the improved STET result, (e) improved STET results for harmonic-like structure and (f) improved STET results for impulsive-like structure.

To test the noise robustness of the improved STET, a series of noises with different signal-to-noise ratio (SNR) levels from -10 dB to 25 dB are added into the simulated signal. For comparison, we list the Rényi entropies of different techniques in Fig. 8. The Rényi entropy increases with increasing input noise, which indicates that the input noise reduces the energy concentration of all TFRs. As shown in Fig. 8, the Rényi entropy of the improved STET is the lowest at different noise levels, which indicates that the energy concentration and noise robustness of the improved STET are the best among all the compared techniques.

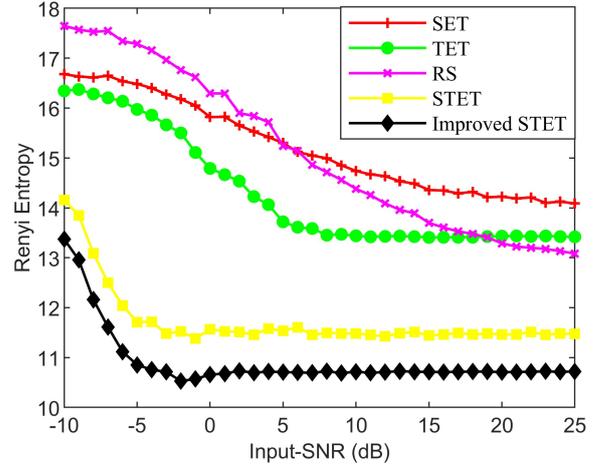

Fig. 8. The Rényi entropies of the TFRs generated by different techniques at different noise levels.

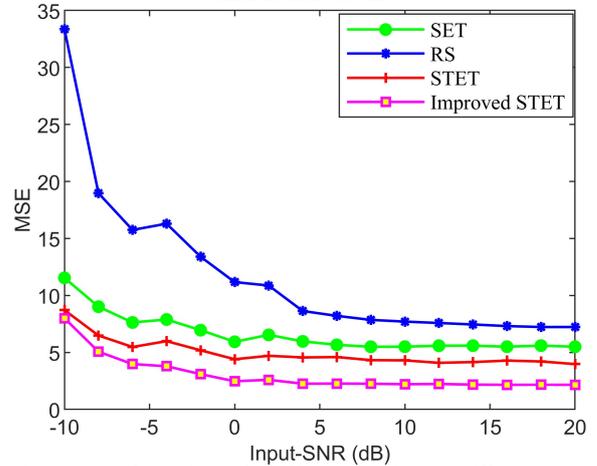

Fig. 9. MSE of IF estimate for different techniques at different SNRs.

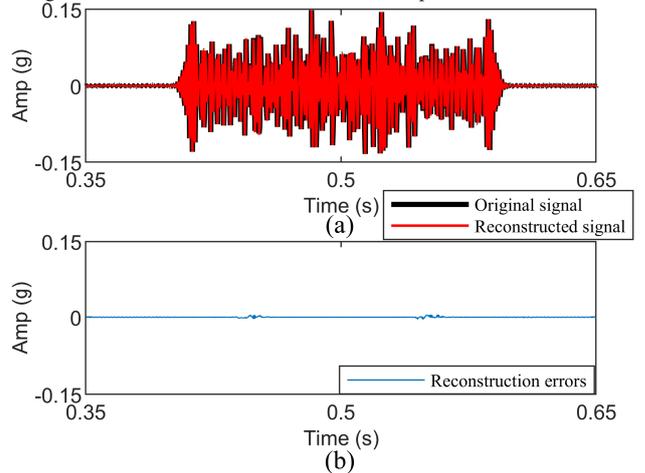

Fig. 10 (a) Reconstruction result of the improved STET and (b) reconstruction errors.

The accuracy of IF estimate is an important factor that affects the energy concentration of the proposed technique. For small chirp signals, most techniques can achieve accurate IF estimate. To demonstrate the performance of the proposed technique, Fig. 9 shows the mean square error (MSE) of the IF estimate and the true IF of different techniques for large chirp signal (chirp

rate=$10^5$ Hz/s ) at different SNRs. The smaller the MSE, the more accurate the IF estimate. It is shown that the improved STET achieves the minimum MSE at different SNRs, which indicates that the accuracy of its IF estimate is higher than other techniques. To evaluate the reconstruction performance of the proposed technique, Fig. 10 shows the reconstruction results of the improved STET. It can be observed that the reconstructed signal is highly consistent with the original signal, and the reconstruction errors is very small compared to the original signal, which indicates that the proposed technique has good reversibility.

## V. EXPERIMENTAL VALIDATION

In this section, we use three real-world signals to illustrate the effectiveness of the improved STET.

### A. Rotating Machinery Rub-impact Fault Vibration Signal Analysis

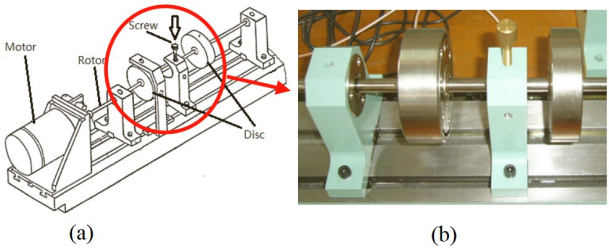

Fig. 11. (a) Mechanical structure schematic of the test rig and (b) a schematic diagram of the test rig.

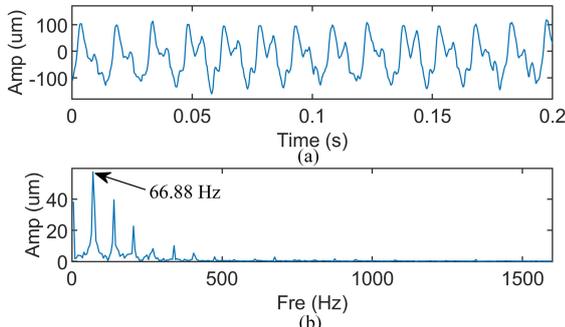

Fig. 12. (a) The waveform of the vibration signal and (b) its spectrum.

Rotor systems play an extremely important role in the mechanical fields. To improve efficiency, seal clearances and bearing clearances in high-speed rotating machinery will be designed to be small. However, a problem caused by too small a clearance is that it may cause friction between the rotor and the stationary elements. Therefore, we use the proposed technique to analyze a rub-impact fault of a rotating machine [22]. The structure of the machine is shown in Fig. 11. Fig. 11(b) shows the way to achieve rub-impact by using the rubbing screw. When it is necessary to do a rub-impact experiment, the screw is gently tightened to make contact with the rotating shaft to generate a single-point rubbing. The rotor vibration is measured by a displacement transducer, where the sampling frequency is 25600 Hz and the sampling time is 0.2 s. The rotational speed of the machine is 4013 rpm and the rotational frequency is 66.88 Hz.

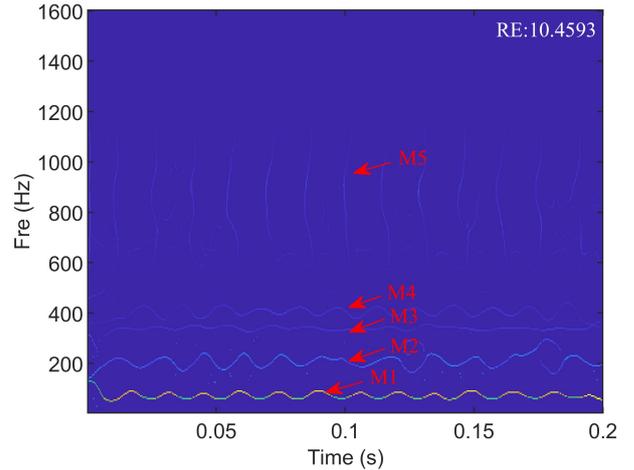

Fig. 13. Improved STET result.

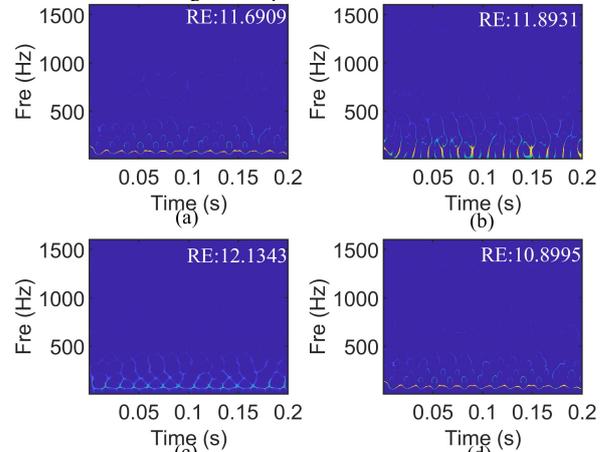

Fig. 14. (a) SET result, (b) TET result, (c) RS result and (d) STET result.

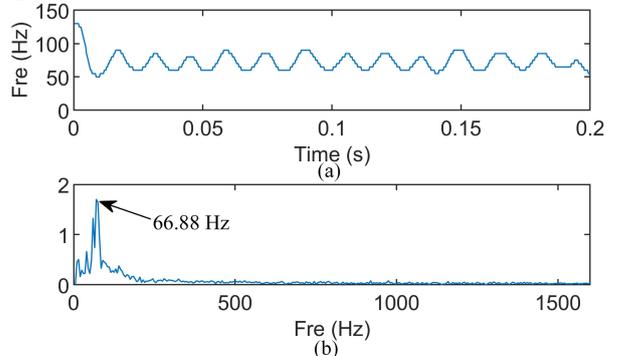

Fig. 15. (a) Extracted IF trajectory of M1 and (b) its spectrum.

The signal waveform and spectrum are shown in Fig. 12. The spectrum shows the most prominent component is 66.88 Hz, which is the rotational frequency. However, it is impossible to accurately determine the fault from the waveform and spectrum. Fig. 13 shows the improved STET result, and we can see that several components are clearly represented. Among them, there are periodic oscillations from M1 to M4, which are due to transient deceleration and acceleration effects as the rotor passes through the fault. In addition, the improved STET captures some impulses (denoted by M5) at high frequency. For comparison, Fig. 14 shows the TFRs of other advanced techniques, and the white numbers in the figures are the Rényi entropy. Compared with other techniques, the improved STET can obtain a higher resolution TFR and has the smallest Rényi entropy, which indicates that the energy concentration of the improved STET are better than other techniques.

The instantaneous frequency at the fundamental frequency of the vibration signal of rotating machinery represents the instantaneous speed of the rotor, so its accurate estimation is the basis of fault diagnosis of rotating machinery. Fig. 15 shows the M1 component extracted using the ridge extraction algorithm and the spectrum of M1. The spectrum shows that the most prominent component in M1 is consistent with the rotational frequency, which indicates that the rotor passes through the fault once per revolution.

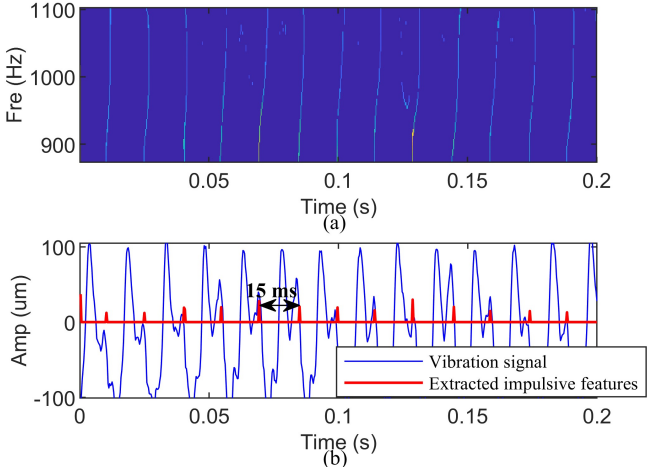

Fig. 16. (a) Improved STET result and (b) extracted impulsive features.

In addition to harmonic oscillation, Fig. 16 shows the impulsive characteristics extracted by the improved STET. It is clear that the improved STET captures 14 fault impulses. The reason why impulses are generated is that when the rotor passes through the fault, it generates amplitude modulation, which eventually leads to transient characteristics at high frequency. Reference [23] shows that the early fault is usually related to the amplitude modulation in the vibration signal, so we should also pay attention to the components at high frequency. We extract impulsive features and plot them with the vibration signal in Fig. 16(b). The result shows that the interval between two consecutive impulses is 15 ms (66.88 Hz), which is the same as the rotational frequency. Therefore, we can conclude that the improved STET can be used for the analysis of rub-impact fault of rotating machinery.

### B. Bearing Fault Diagnosis

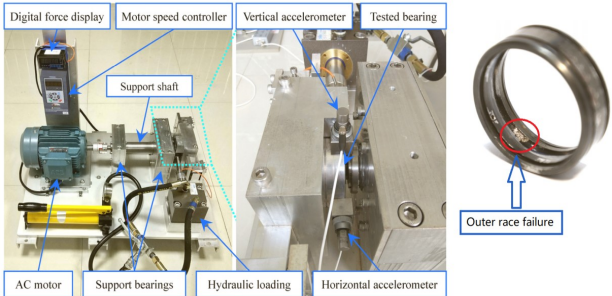

Fig. 17. Schematic diagram of the test rig.

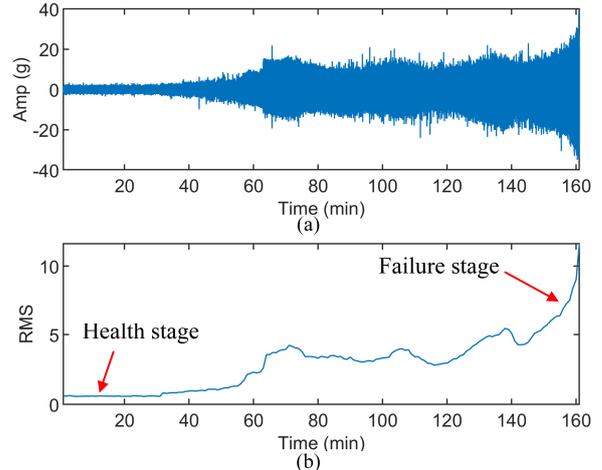

Fig. 18. (a) Signal waveform and (b) RMS.

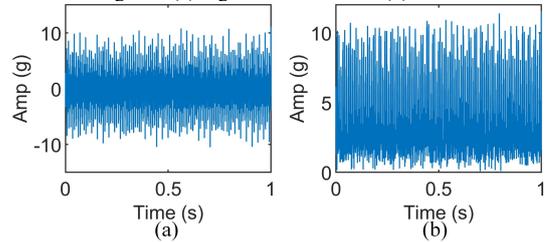

Fig. 19. (a) Signal waveform (b) the envelope of the signal.

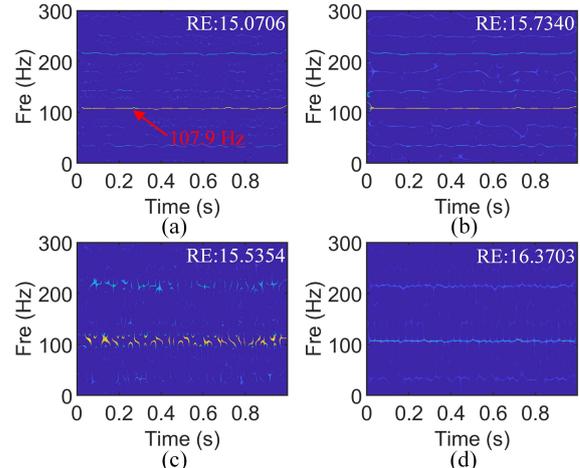

Fig. 20. The TFR of (a) improved STET, (b) SET, (c) TET and (d) RS.

In this section, a bearing vibration signal with an outer race fault is used to verify the effectiveness of the improved STET. Fig. 17 shows the structure diagram of the experimental setup [24]. The tested bearing model is LDK UER204, and the working speed of the tested bearing is 2100 rpm. The sampling rate is 25600 Hz, and the fault characteristic frequency is 107.9 Hz according to the bearing parameters [25].

Fig. 18 shows the signal waveform and its root mean square (RMS) value. In the first 30 minutes, the RMS value remains low and hardly changes. After 30 minutes, a bearing outer race fault occurs and the RMS changes drastically.

We select a segment of the signal in the failure stage for analysis. The fault signal and its envelope are shown in Fig. 19. The results of the different techniques for processing the envelope are listed in Fig. 20. The fault frequency can be observed from the TFRs of the improved STET, SET and RS.

However, the improved STET has the smallest Rényi entropy, which indicates that it has the highest energy concentration, and this is more beneficial for achieving accurate bearing fault diagnosis. Fig. 21 shows the results of different techniques for processing a segment of the fault signal in Fig. 19(a). The improved STET accurately extracts 11 impulses and generates a highly concentrated TFR for each impulse. The time interval between two consecutive impulses is measured to be 9.3 ms (107.9 Hz). This is consistent with the theoretical value, which indicates the occurrence of bearing outer race faults. Furthermore, the improved STET has the smallest Rényi entropy. The resolution of other TFA techniques is not as good as that of the improved STET, which makes it impossible to accurately determine the type of fault. The improved STET can achieve accurate fault diagnosis based on both harmonic and impulsive components of the signal, while the rest of the techniques are unable to extract both harmonic and impulsive components, which highlights the advantages of the proposed technique.

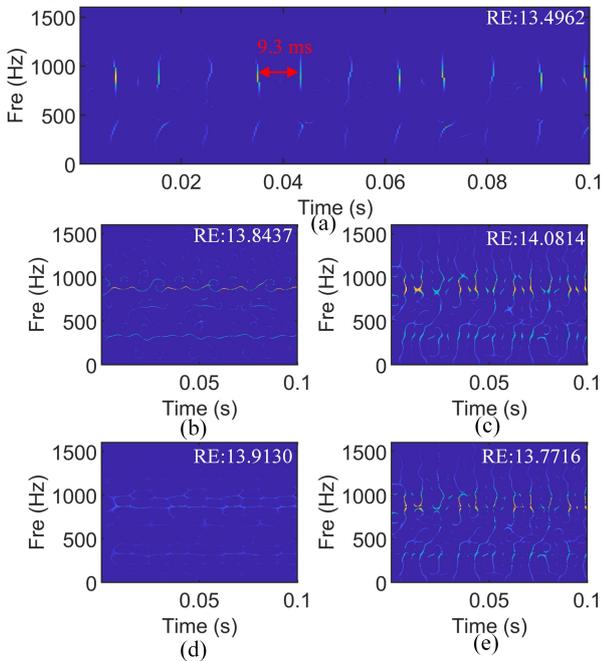

Fig. 21. The TFR of (a) improved STET, (b) SET (c) TET, (d) RS and (e) STET.

### C. Gear Fault Analysis under Time-varying Speed

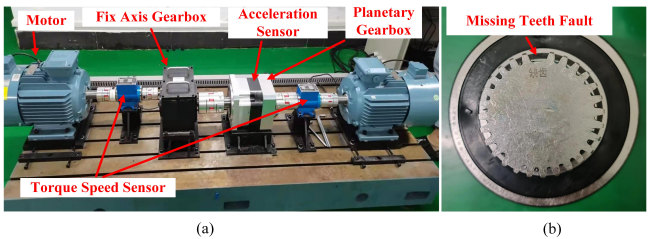

Fig. 22. Schematic diagram of (a) the test rig and (b) a faulty sun gear with missing teeth.

In this section, the validity of the proposed technique is verified using a sun gear signal with missing teeth from the HOUDE METERS HD-CL-05 planetary gearbox test rig. The structure of the test rig and the faulty gear are shown in Fig. 22. The test rig contains a main drive motor, a planetary gearbox, a fixed axis gearbox and a load motor. A B&K4370 accelerometer is installed on the planetary gearbox to acquire condition detection signals at a sampling frequency of 10 kHz. The shaft speed increased from 1200 rpm to 1860 rpm within 3.4s. Table I lists the parameters of the planetary gearbox and the fault characteristic frequency of the faulty gear.

TABLE I
**Planetary gearbox parameters**

| Sun Gear (Number of Teeth) | Planet Gear (Number of Teeth) | Gear ring (Number of Teeth) | Number of Planet Gear |
|---|---|---|---|
| 28 | 28 | 84 | 4 |

| Motor speed | Sun gear rotation frequency ($f_r$) | Sun gear localized fault frequency ($f_c$) | |
|---|---|---|---|
| 1200-1860 rpm | 20-31 Hz | $3f_r$ | |

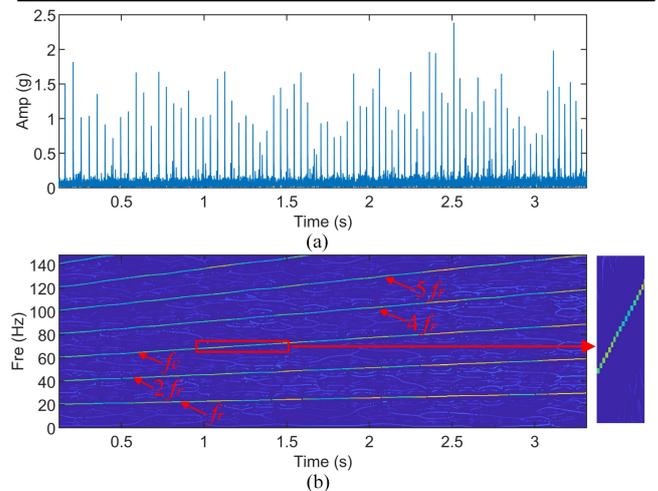

Fig. 23. (a) Envelope of the signal and (b) improved STET result.

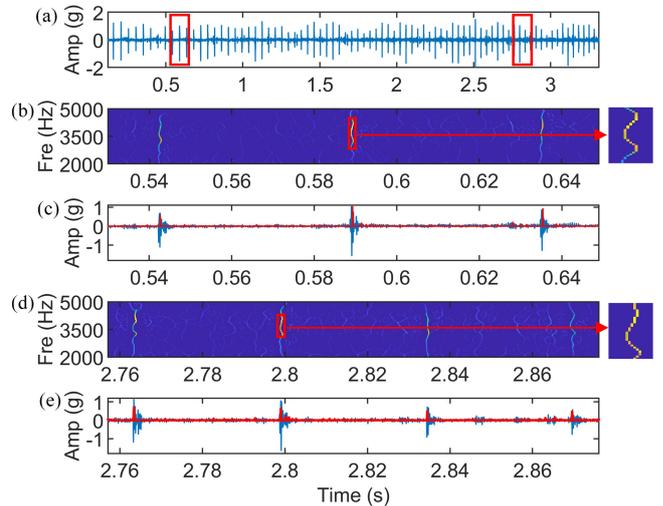

Fig. 24. (a) Signal waveform, (b) improved STET result during 0.53-0.649 s, (c) slice of the improved STET results during 0.53-0.649 s, (d) improved STET result during 2.757-2.876 s and (e) slice of the improved STET results during 2.757-2.876 s.

Fig. 23 shows the envelope of the signal and the TFR of the improved STET. It is shown that the improved STET accurately extracts the rotational frequency of the sun gear and its higher harmonics as well as the fault frequency. The zoom of the improved STET result shows that it generates an energy concentrated TFR for the fault frequency, which can be used to determine the fault. Fig. 24 shows the waveform of the signal. Two signal segments are selected and analyzed using the improved STET and the results are shown in Figs. 24(b)-(e). The improved STET generates energy concentrated TFR for each impulse. Then, the impulsive features are extracted and compared with the signal as shown in Figs. 24(c) and (e). The results show that the improved STET can accurately extract faulty impulses from the signal under time-varying speed. This example shows that the improved STET can extract the harmonic and impulsive components of the faulty gear signal for fault diagnosis, which further demonstrates the practicality of the proposed technique in complex work.

## VI. CONCLUSION

In this paper, we propose a novel TF postprocessing technique for analyzing non-stationary signals from rotating machinery. The proposed technique can adaptively extract the time-frequency ridges of different components of the signal according to the chirp rate. Existing TF postprocessing techniques all have implicit assumptions on the analyzed signal, while the main advantage of the proposed technique is that it breaks this limitation. This enables the proposed technique to simultaneously provide energy concentrated TFR for both harmonic and impulsive components of a signal, while allowing the reconstruction of the interested components. The effectiveness of the proposed technique is verified using two numerical signals and three experimental signals. All results show that the proposed technique produces TFR with higher energy concentration and better noise robustness than other advanced TFA techniques.

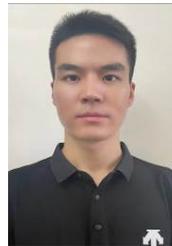

**Yunlong Ma** was born in Linyi, China in 1997. He graduated from Shandong Management University in 2018.

He is currently a master student of Electrical Engineering with the University of Jinan, Jinan, China. His current research interests include time-frequency analysis, machinery condition monitoring, and fault diagnosis.

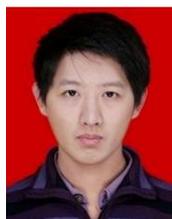

**Gang Yu** was born in Zhangqiu, China, in 1987. He received the B.Eng. degree in mechanical engineering from Qingdao University, Qingdao, China, and the Ph.D. degree in mechanical engineering from Shandong University, Jinan, China, in 2010 and 2016, respectively.

He is currently a Lecturer of Mechanical Engineering with the University of Jinan, Jinan, China. His current research interests include time-frequency analysis, machinery condition monitoring, and fault diagnosis.

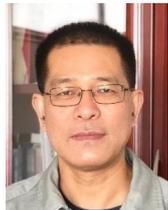

**Tianran Lin** received the B. Eng. degree in mechanical engineering from Jiangnan University, Wuxi, China, in 1987, and the M. Eng. and Ph.D degrees in mechanical engineering from the University of Western Australia, Perth, Australia, in 2001 and 2006, respectively.

He is currently a 'Taishan Scholar' distinguished Professor in the School of Mechanical and Automatic Engineering at Qingdao University of Technology, Qingdao, China. His research interests include noise and vibration analysis and control, signal processing, machine fault diagnosis and artificial intelligent. Pro. Lin is a fellow of the International Society of Engineering Asset Management.

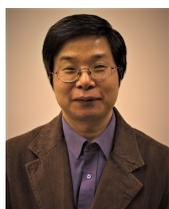

**Qingtang Jiang** received the B.S. and M.S. degrees from Hangzhou University, Hangzhou, China, in 1986 and 1989, respectively, and the Ph.D. degree from Peking University, Beijing, China, in 1992, all in mathematics.

He was with Peking University from 1992 to 1995. He was an NSTB Postdoctoral Fellow and then a Research Fellow at the National University of Singapore from 1995 to 1999. Before he joined the University of Missouri-St. Louis, in 2002, he held visiting positions at the University of Alberta, Canada, and West Virginia University, Morgantown, USA. He is currently a Professor in the Department of Math and Statistics, University of Missouri-St. Louis, St. Louis，USA. His current research interests include time-frequency analysis, wavelet analysis, signal processing, and image restoration.